\documentclass[twocolumn, showpacs,preprintnumbers,amsmath,amssymb]{revtex4}

\usepackage{graphicx}
\usepackage{dcolumn}
\usepackage{bm}

\begin{document}

\title{Effect of hybridization on the magnetic properties of  correlated two-band metals}
\author{C.M. \surname {Chaves}${^1}$}
\email{cmch@cbpf.br}
\author{ A. \surname{Troper}${^{1,2}}$}
\affiliation{{$^1$}Centro Brasileiro de Pesquisas F\'\i sica, Rua Xavier Sigaud 150, Rio de Janeiro, 22290-180, RJ, Brazil}

\affiliation{${^2}$Universidade do Estado do Rio de Janeiro, Rua S. Francisco Xavier 524,20550-013,Rio de Janeiro RJ, Brazil}

\date{\today }

\begin{abstract}
The magnetic properties of transition-like metals are discussed within the single site approximation, which is a picture  to take into account electron correlations. The metal is described by two hybridized bands one of which includes Coulomb correlation. The presented results indicate that ferromagnetism arises for adequate values of hybridization ($V$), correlation ($U$) and occupation number($n_{\sigma}$). Some similarities with Dynamical Mean-Field Theory (DMFT) are indicated. 
\end{abstract}

\pacs{71.10.-w, 71.10.Fd}
\maketitle
\section{Introduction}
\label{sec:intro}

Recently\cite{sch,bat}, the conventional view of the origin of ferromagnetism in metals has been under criticism. Traditional mean-field calculations favor ferromagnetism but corrections tend to reduce the range of validity of that ground state \cite{voll2}.

Several recent works addressed the issue of ferromagnetism in metals, going beyond the Stoner model : Vollhardt \textit{ et al}\cite{voll} furnish evidence of the stability of itinerant ferromagnetism in the one-band Hubbard model(HM)at electronic densities not too close to half-filling and large enough $U$. 

Nolting and Borgiel\cite{nol} use, for the one-band Hubbard model(HM), a spectral density approach(SDA), a two-pole ansatz that is equivalent to the Roth method\cite{roth2}, containing some free parameters to be determined by fitting some spectral moments. Ferro-and antiferromagnetic ~states are obtained. Ferromagnetic solution occurs only if the occupation number $n$  exceeds a critical value and $U$ exceeds a minimum value. The ferromagnetic order arises due to a shift of the $\uparrow$ and $\downarrow$ quasiparticle subbands.

Jarrel\cite{jarrel} uses a self-consistent quantum Monte-Carlo method for the $d=\infty$ one-band HM but did not find a ferromagnetic state for any filling, temperature or correlation.

Herrmann and Nolting\cite{herr}, also in the one-band HM, use the spectral density approach (SDA), and a modified alloy analogy(MAA) method. In the latter, the atomic levels of the fictitious alloy are the ones found with the ~SDA. In both, a ferromagnetic state develops, although in the MAA it happens in a rather restricted region of occupation $n$. They point out that an imaginary part in the self-energy, present in the MAA method but not in the two-pole ansatz, causes important differences in the magnetic order of the model. 
An improved version of the AA which incorporates the intersite magnetic correlations\cite{bh} seems to indicate a ferromagnetic instability for large $U$ and $n\sim 0.6$. 

Schwieger and Nolting \cite{sch} consider a two-band HM, as we do, but for $d\rightarrow \infty$ and use both the SDA and the MAA methods and show that interband particle fluctuations cause a spin dependent band shift and a spin dependent broadeninig of the Hubbard bands. This shift stabilizes and the broadening destabilizes ferromagnetism. For $U=5$ the critical temperature $T_c$ is plotted versus hybridization; strong fluctuations between the bands tend to suppress ferromagnetic order.

We use instead a single-site approximation (SSA)\cite{roth1}. In this formulation, only one site, the origin say, exhibits the full Coulomb interaction; the others are subject to a local field: the spin and energy dependent self-energy $\Sigma^{\sigma}$. $\Sigma^{\sigma}$  is self-consistently determined by imposing the vanishing of the scattering matrix, a condition formulated in the Coherent Potential Approximation(CPA) for alloys and impurity problems, to restore translation invariance. We will be working at temperature $T=0$.

The metal is described by two non-degenerate bands, $a$ and $b$; $a$ is a Hubbard-like narrow band with in-site interaction $U$, hybridized($V_{ab}$) with the second one, $b$, a broader uncorrelated band. 

\section{Theoretical Model: The Self-Consistent SSA}
\label{sec:ssa}

The starting Hamiltonian in this work is

\begin{eqnarray}
\label{H}
\mathcal{H}&=&\sum_{i,j,\sigma }t_{ij}^{a}a_{i\sigma }^{+}a_{j\sigma
}+\sum_{i,j,\sigma }t_{ij}^{b}b_{i\sigma }^{+}b_{j\sigma
}+\sum_{i}Un_{i\uparrow }^{(a)}n_{i\downarrow }^{(a)} \nonumber\\
&+&\sum_{i,j,\sigma }(V_{ab}b_{i\sigma }^{+}a_{j\sigma }+V_{ba}^{+}a_{i\sigma}^{+}b_{j\sigma })~,
\end{eqnarray}
where  $n_{i\sigma }^{a}=a_{i\sigma }^{+}a_{i\sigma }$ and $\sigma$ denotes spin. $t_{ij}^{a,b}$ is the tunneling amplitude between neighboring sites $i$ and $j$ , in each band and $V_{ab}$ the hybridization. In the single-site approximation \cite{roth1} one adopts the following  effective Hamiltonian:
\begin{eqnarray}
\label{Heff}
 \mathcal{H}_{eff}&=&\sum_{i,j,\sigma }t_{ij}^{a}a_{i\sigma }^{+}a_{j\sigma
}+\sum_{i,j,\sigma }t_{ij}^{b}b_{i\sigma }^{+}b_{j\sigma
}+\sum_{i\not=0,\sigma}n_{i,\sigma}^{a}\Sigma^{\sigma}\nonumber\\
&+&Un^{a}_{0\uparrow }n^{a}_{0\downarrow} 
+\sum_{i,j,\sigma } (V_{ab}b_{i\sigma }^{+}a_{j\sigma }
+V_{ba}^{*}a_{i\sigma}^{+}b_{j\sigma })
\end{eqnarray}
The  method then replaces a translationally invariant problem, as defined by (\ref{H}), by an impurity problem where only the origin incorporates the Coulomb interaction, the other sites being acted by the local (k-independent) field $\Sigma^\sigma$. But the effective Hamiltonian (\ref{Heff}), describes an `impurity problem' in presence of Coulomb intra-atomic term and we have to resort to some approximation. 

We use the Green function method \cite{zuba}; after some algebra one obtains for the $a$-band Green function
\begin{eqnarray}
\label{gaa}
 G^{aa}_{kk'\sigma}(w)&=&{\frac{\delta_{kk'}}{w-\tilde{\epsilon}^a_k-\Sigma^\sigma}}\nonumber\\
&+&{\frac{1}{w-\tilde{\epsilon}^a_k-\Sigma^\sigma}}T^\sigma(w,\Sigma^\sigma){\frac{1}{w-\tilde{\epsilon}^a_{k'}-\Sigma^\sigma}} 
\end{eqnarray}
In this expression 
\begin{equation}
\tilde {\epsilon}^{a}_{k}=\epsilon^{a}_k +{\frac{V_{ab}^2(k)}{w-{\epsilon}^b_k }}, 
\label{etild}
\end{equation}
is  the recursion relation of the $a$ band including hybridization and 
\begin{equation}
\epsilon^{a}_{k}={\frac{t_{a}(cos(k_xa)+\cos(k_ya)+\cos(k_za))}{A}}, 
\label{ed}
\end{equation}
is the respective recursion relation of the bare band. In this paper we use $t_a =1$ and $A=3$, in arbitrary energy units. All energy magnitudes are taken in units of $t_a$, making them dimensionless. The bare $a$ band width is then $W=2$. For simplicity we use $\epsilon^{b}_{k}=\alpha \epsilon^{a}_{k}+\epsilon_s$ (homothetic bands), for the bare $b$ band. $\epsilon_s$ is the shift between the center of the bands. From now on we take  $k_ia \rightarrow k_{i}, i=x,y,z$ and $V_{ab}=V_{ba}\equiv V=$constant independent of $k_i$.
The scattering $T^\sigma$-matrix in Eq.(\ref{gaa}) is given by
\begin{widetext}
\begin{equation}
T^{\sigma}(w,\Sigma^\sigma)={\frac{U<n^a_{0-\sigma>} -\Sigma^\sigma(w)+\Sigma^\sigma(w)(U-\Sigma^\sigma(w))F^\sigma(w)}{[1-(U-\Sigma^\sigma(w))F^\sigma(w)][1+\Sigma^\sigma(w) F^\sigma(w)]}}. 
\label{T}
\end{equation}
\end{widetext}
where
\begin{equation}
F^{\sigma}(w)=\sum_k {\frac{1}{w-\tilde{\epsilon}^a_k-\Sigma^\sigma(w)}}. 
\label{F}
\end{equation}
The self-energy $\Sigma^\sigma$ is complex and spin dependent-thus giving a spin dependent band shift.
The vanishing of the T-matrix gives a self-consistent equation for the self-energy:
\begin{equation}
\label{elivre}
\Sigma^\sigma(w)= U<n^a_{0-\sigma}>+\Sigma^\sigma(w)(U-\Sigma^\sigma(w))F^\sigma(w).
\end{equation}
It is important to stress that Eq.(\ref{elivre}) results from a \textit {configuration} average caracteristic of a CPA approach and not from a dynamical constraint.

The Green function $G^{aa}_{k\sigma}$ then becomes
\begin{eqnarray}
\label{ga} 
 G^{aa}_{k\sigma}(w)={\frac{1}{w-\tilde{\epsilon}^a_k-\Sigma^\sigma(w)}}
\end{eqnarray}
while $G^{bb}_{k\sigma}$ is
\begin{eqnarray}
\label{gbb}
 G^{bb}_{k\sigma}(w)= G^{b}_{k}(w) +V^2G^{b}_{k}(w)G^{aa}_{k\sigma}(w)G^{b}_{k}(w) 
\end{eqnarray}
with 
\begin{eqnarray}
\label{gb}
 G^{b}_{k}(w)={\frac{1}{w-\epsilon^b_k}},
\end{eqnarray}
the Green function of the bare b band. 

The procedure presented here has some  similarity to the one used in the Dynamical Mean-Field Theory (DMFT)\cite{gabi}in the following sense: here the original lattice model with correlation in every site is replaced by an effective one,  where the correlation is present only at the origin ('the impurity') but at the same dimension $d=3$. 

\section{Numerical Results and Conclusions}
\label{sec:results}

\begin{figure}
\vspace {0.1cm}
\resizebox{80mm}{!}
{\includegraphics{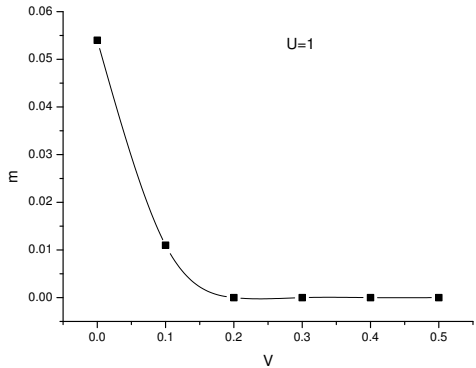}}
\caption{ (Color online)Magnetization $m=<n_{\uparrow}^{a}>-<n_{\downarrow}^{a}>$ of the correlated a-band versus $V$ in a regime $U/W<1$, namely,
 $U=1$.}
\label{mv1}
\end{figure}

\begin{figure}
\vspace {0.1cm}
\resizebox{80mm}{!}
{\includegraphics{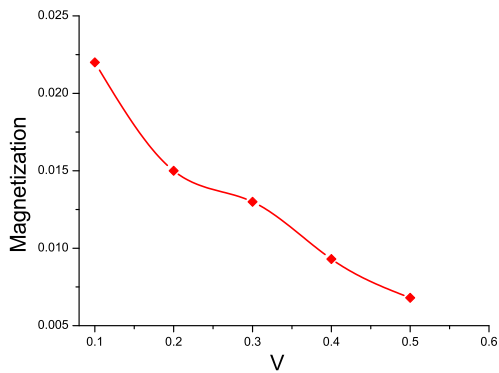}}
\caption{ (Color online)Magnetization $m=<n_{\uparrow}^{a}>-<n_{\downarrow}^{a}>$ of the correlated a-band versus $V$ in a regime $U/W\sim 1$, namely $U=2$.}
\label{mv2}
\end{figure}

\begin{figure}
\vspace {0.1cm}
\resizebox{80mm}{!}
{\includegraphics{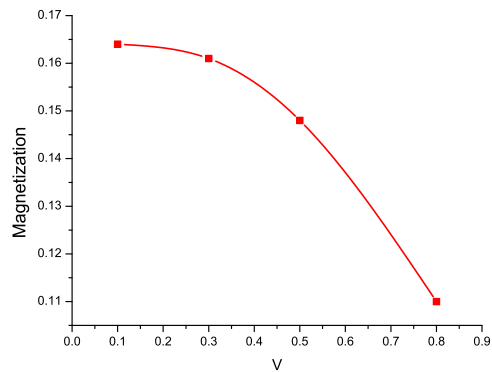}}
\caption{ (Color online)Magnetization $m=<n_{\uparrow}^{a}>-<n_{\downarrow}^{a}>$ of the correlated a-band versus $V$ in a regime $U/W>1$, namely $U=5$.}
\label{mv3}
\end{figure}

In the numerical calculations one chooses the total number of electrons per site as being about $n=2.2$, a little more than half-filling. We start with $<(n^{a}_{\uparrow})^0>=0.52$ and $<(n^{a}_{\downarrow})^0>=0.45$ in eq(\ref{elivre}), searching for a ferromagnetic solution in a less than half filled a-band. The dynamics generated by $V$ and $U$, then, redistributes the $a$ and  the $b$ band electrons, mixing the up and down states eventually producing a magnetization. In ref.\cite{sch} however, the density of the correlated band, rather than $n$, is kept fixed. For comparison, in figs (\ref{mv1}) and (\ref{mv2}), we took $n=2.2$, $\alpha=1.5$ and $\epsilon_s=1.0$ while in fig (\ref{mv3}), $\alpha=2.5$, $n=2.0$ e $\epsilon_s=1.0$.

We found that, for a certain range of parameters $U$ and $V$, a ferromagnetic state (FS) develops. We then investigate the effect of $V$ in this state, for a given $U$ in the weak ($U<W$), in the intermediate ($U\sim W$) and in the strong coupling ($U>W$) limits. 
In fig (\ref{mv1}) we display the magnetization $m$ versus $V$ for $U=1$. For $V \geq 0.3$ the magnetization is already very small or zero. For a given U, the effect of the hybridization is such as to reduce or suppress the magnetization. 

In the intermediate limit, e.g., $U=2$, $U/W \sim 1$, as shown in fig(\ref{mv2}), ferromagnetism  is also observed ; again increasing $V$ the metal tends to a non-magnetic state.

In fig(\ref{mv3}) we exhibit the magnetization for a typical strong coupling situation, namely, $U=5$. We notice that the decrease of the magnetization with $V$ is now slower than for smaller $U$. 

We have shown that the present method is computationally feasible, producing reliable and physically sensible results compatibles with the existing literature on this important subject, opening new insights concerning the possible $U/W$ regimes. Further work, for $T>0$, and extending the model to include a two sublattice system in order to describe possible antiferromagnetic states, is in progress.

\subsection*{Acknowledgments}

CMC and AT acknowledge the support from the brazilian agencies $PCI/MCT$ and $CNP_q$.

\end{document}